\documentstyle[epsfig,12pt]{article}
\begin{document}

\title{The topology of cross-border exposures: beyond the minimal spanning
tree approach }
\author{Alessandro Spelta \thanks{%
corresponding author: alessandro.spelta01@ateneopv.it} \ and Tanya Ara\'{u}%
jo \thanks{%
tanya@iseg.utl.pt} \\
%EndAName
University of Pavia, Italy\\
ISEG - Technical University of Lisbon (TULisbon) and\\
Research Unit on Complexity in Economics (UECE), Portugal}
\date{}
\maketitle

\begin{abstract}
The recent financial crisis has stressed the need to understand financial
systems as networks of interdependent countries, where cross-border
financial linkages play the fundamental role. It has also been emphasized
that the relevance of these networks relies on the representation of changes
follow-on the occurrence of stress events. Here, from series of interbank liabilities
and claims over different time periods, we have developed networks of
positions (net claims) between countries. Besides the Minimal Spanning Tree
analysis of the time-constrained networks, a coefficient of residuality is
defined to capture the structural evolution of the network of cross-border
financial linkages. Because some structural changes seem to be related to
the role that countries play in the financial context, networks of debtor
and creditor countries are also developed. Empirical results allows to
relate the network structure that emerges in the last years to the globally
turbulent period that has characterized financial systems since the latest
nineties. The residuality coefficient highlights an important modification
acting in the financial linkages across countries in the period 1997-2011,
and situates the recent financial crises as replica of a larger structural
change going on since 1997.
\end{abstract}

\section{Introduction}

Despite the conventional wisdom that national economies are interdependent
there is less evidence on the contribution of economic crises to the
reinforcement of cross-border financial interdependencies. The recent
financial crisis has stressed the need to understand financial systems as
networks of countries where cross-border financial linkages play the
fundamental role. Furthermore, it has become clear that the relevance of
these networks relies on the representation of changes follow-on the
occurrence of stress events.

The adoption of an evolving network approach is recommended not only because
of the proper emphasis on the financial interdependencies but also due to
the possibility of describing how the structure of these interdependencies
evolves in time. In so doing, we are able to address the role that an
existing network structure plays in the spread of shocks and conversely, the
effectiveness of stress events and their impact on the structure of the
network.

In August 2007, the crisis of the subprime mortgage industry stormed the
financial systems of several countries. As a response, hundreds of billions
of dollars were injected by the authorities into the market. Nevertheless,
this was not enough to avoid the second wave of a financial crisis.

Since the end of 2009, the down grading of Greek debt has been followed by
several debt crises in the EU member states. As many of these countries cope
with a decline in growth prospects and an increase in debts loads, the
spread of the shocks seems to lead the EU to an unsustainable macro
situation [1].

Deriving macro (global) situations from micro (local) scenarios has been a
recurrent topic in economics. The way to link the macro and the micro levels
hinges on graph theory, which has been recently introduced in economics to
study formally the generation and stability of economic interactions among
agents ([2] [3]). Even though basic geometric and topologic structures have
already been used by Helpman and Krugman [4] to explain international trade
relations, the knowledge on the complex structure of financial networks is
far from being sufficient to realize the potential of the topological
approaches.

Several authors have reported the opportuneness of describing financial
linkages through a network approach. Some papers have favored the study of
interdependencies between credit banks [5], or focused on the analysis of
shocks storming the financial systems of several countries [6]. The
topological properties of national interbank markets have also been studied
by Soramaky and co-workers [7] as they analyzed the network topology of the
interbank payments transferred between commercial banks over the Fedwire
Funds Services [8]. In the same context, Fuijwara and collaborators explored
the credit relationships between commercial banks and a set of Japanese
companies [9].

The adoption of a network approach on empirical data has also been conducted
in the characterization of the Italian [10] and the Austrian [11] interbank
markets. More recently, Kubelec and collaborators [12] as well as Garratt\
and co-workers [13] used the Bank for International Settlements (BIS)
consolidated banking statistics to develop cross-border banking networks.
McGuire and co-workers \ [14] studied the international banking system while
focussing in the cross-border financial trade. A similar approach was
adopted by Minoiu and co-workers [15] in their study of global banking
networks. For several extreme phenomena taking place at the global economic
level, they found enough evidence of incoming instabilities in the network
regime.

Starting from a geometrical setting we also followed a topological approach.
Topological coefficients have been the object of growing attention ever
since some network regimes were identified as the underlying structures of
important phenomena found in many different fields. However, as most of
these coefficients apply to graph structures that are connected and sparse,
when analysing systems whose topological signature is a complete
(fully-connected) network, there is a need to find out the corresponding
representation of the system where sparseness replaces completeness in a
suitable way. It has been often accomplished ([16], [17], [18 ] and [19])
through the construction of a Minimal Spanning Tree (MST).

In the present paper we have explored an equivalent approach to construct a
sparse and connected network of financial linkages across countries. From
time series of interbank liabilities and claims conducted through the
international banking system, we have developed networks of countries'
positions (net claims). The networks are built for the last 28 years and for
two chronologically successive branches of 14 years.

In this context, besides the MST analysis of the cross-country networks, a
coefficient of redundancy is defined to capture the structural changes
occurring on the network along the last 28 years.

Because some structural changes seem to be related to the role that
countries play in the financial context, networks of debtor and creditor
countries are also developed. Empirical results allows to relate the network
structure that emerges in the last years to the globally turbulent period
that has characterized financial systems since the latest nineties. The
residuality coefficient highlights an important modification acting in the
financial linkages in the period 1997-2011, and situates the turbulent
period that has been characterizing the global financial system since the
Summer 2007 as replica of a larger structural change going on for a decade.
This occurs in such a systematic fashion that it may not be a statistical
accident.

The paper is organized as follows: Section 2 briefly presents the set of
empirical data we work with. Section 3 is targeted at presenting the method
and the first results obtained from its application. The main contributions
of the paper are presented in Sections 4 and 5: the coefficient of
residuality and its application to the evaluation of the structural
evolution of some specific networks as those that result from the
identification of debtor and creditor countries. The paper ends with
appropriate conclusions and outline of future work.

\section{The data}

The Bank for International Settlements (BIS) locational banking statistics
(IBLR), including international claims and liabilities of reporting banks by
country of residence and for the reporting area as a whole, provides a
plentiful data set of aggregate cross-border exposures for a set of
reporting and non-reporting countries all over the world.

The locational statistics were originally intended to complement monetary
and credit aggregates, being consistent with both the national balances of
payments and the systems of national accounts. For a subset of 57 reporting
countries, the BIS publishes bank claims\ ($C$) \ and liabilities ($L$) on
all other countries in a quarterly basis.

Our approach is applied to a subset of 24 reporting countries (see Table 1)%
\footnote{%
The first data collection date back to 1977 but a significant amount of
statistics can be obtained only from the eighties. The statistics prior to
1983, includes just fifteen countries.} and from this data source, bilateral
exposures between each pair of countries are used to compute the \textbf{%
position} of each country \textit{vis-a-vis} the rest of reporting countries.

The position $p_{i,t}$ of country $i$ in quarter $t$ is obtained subtracting
liabilities ($L_{i,t}$) from claims ($C_{i,t}$)

\begin{equation}
p_{i,t}=C_{i,t}-L_{i,t}
\end{equation}

Thus, the position of each country corresponds to its net claims. From the
countries' positions we are able to classify countries as creditor or debtor
countries accordingly to the sign of $p_{i,t}$.\emph{\ }If $p_{i,t}>0$ then
country $i$ is a creditor but if it has a greater amount of liabilities than
claims the country is a debtor, since $p_{i,t}<0$. Later in the paper
(Section 5) we shall address that issue while applying the classification to
further characterize the reporting countries.

\begin{center}
\begin{tabular}{|c|c|}
\hline
1) AT: Austria & 13) IT: Italy \\ \hline
2) BS: Bahamas & 14) JP: Japan \\ \hline
3) BH: Bahrain & 15) LU: Luxemburg \\ \hline
4) BE: Belgium & 16) NL: Netherlands \\ \hline
5) CA: Canada & 17) AN: Netherlands Antilles \\ \hline
6) KY: Cayman Islands & 18) NO: Norway \\ \hline
7) DK: Denmanrk & 19) SG: Singapore \\ \hline
8) FI: Finland & 20) ES: Spain \\ \hline
9) FR: France & 21) SE: Sweden \\ \hline
10) DE: Germany & 22) CH: Switzerland \\ \hline
11)HK: Hong Kong & 23) GB: United Kingdom \\ \hline
12) IE: Ireland & 24) US: United States \\ \hline
\end{tabular}

Table 1: Reporting Countries
\end{center}

Since the BIS locational banking statistics capture the net flows of
financial capital between any two countries channeled through the banking
system [20], this data set is an appropriate source to the empirical study
of temporal patterns arising from financial linkages across countries.

\section{The Method}

Cross-correlation based distances, as applied in reference [16] to the study
of stock market structure have been used in the analysis and reconstruction
of geometric spaces in many different fields. The quantity

\begin{equation}
d_{kl}=\sqrt{2\left( 1-C_{kl}\right) }  \label{I.1}
\end{equation}

where $C_{kl}$ is the correlation coefficient of two time series $%
\overrightarrow{s}(k)$ and $\overrightarrow{s}(l)$ computed along a given
time window
\begin{equation}
C_{kl}=\frac{\left\langle \overrightarrow{s}(k)\overrightarrow{s}%
(l)\right\rangle -\left\langle \overrightarrow{s}(k)\right\rangle
\left\langle \overrightarrow{s}(l)\right\rangle }{\sqrt{\left( \left\langle
\overrightarrow{s}^{2}(k)\right\rangle -\left\langle \overrightarrow{s}%
(k)\right\rangle ^{2}\right) \left( \left\langle \overrightarrow{s}%
^{2}(l)\right\rangle -\left\langle \overrightarrow{s}(l)\right\rangle
^{2}\right) }}  \label{I.2}
\end{equation}%
has been shown [17] to satisfy all the metric axioms.

\subsection{The Metric}

Using the BIS time series with claims and liabilities and after computing
the quarterly based position of each country ($\overrightarrow{p}(i)$)
\textit{vis-a-vis} the other reporting countries, we define a normalized
vector

\begin{equation}
\overrightarrow{\rho }(i)=\frac{\overrightarrow{p}(i)-\left\langle
\overrightarrow{p}(i)\right\rangle }{\sqrt{n\left( \left\langle
\overrightarrow{p}^{2}(i)\right\rangle -\left\langle \overrightarrow{p}%
(i)\right\rangle ^{2}\right) }}  \label{2.2}
\end{equation}%
$n$ being the number of components (number of time labels) in the vectors $%
\overrightarrow{p}(i)$. With this vector one defines the \textit{distance}
between the countries $i$ and $j$ by the Euclidean distance of the
normalized vectors,

\begin{equation}
d_{ij}=\sqrt{2\left( 1-C_{ij}\right) }=\left\Vert \overrightarrow{\rho }(k)-%
\overrightarrow{\rho }(l)\right\Vert
\end{equation}

where $C_{ij}$ is the correlation coefficient of the positions $%
\overrightarrow{p}(i)$ and $\overrightarrow{p}(j)$, respectively of
countries $i$ and $j$ computed along a time window of $n$ (quarterly)
observations.

Having computed the matrix of distances ($D$) for the set of 24 reporting
countries ($N=24$) on a time interval of 110 quarters ($n=110$), we are able
to provide three pictorial representations of this set, as Figure 1 shows.

Although the first plot presented in Figure 1(a) graphically informs about
the distribution of the distances between countries, the fully-connected
nature of this set does not help to find out whether there is a dominant
pattern taking place. Therefore, the first step should be targeted at
obtaining a sparse representation of the set of distances, with the \textit{%
degree of sparseness} generated by the set of distances itself, instead of
an \textit{a priori} specification. At the same time, when looking for a
suitable degree of sparseness, one must avoid disconnected graphs. To this
end, we construct the corresponding MST.

\subsection{From complete networks to sparse graphs: the minimal spanning
tree approach}

From the $NxN$ distance matrix $D$, a hierarchical clustering is then
performed using the \textit{nearest neighbor} method. Initially $N$ clusters
corresponding to the $N$ countries are considered. Then, at each step, two
clusters $c_{i}$ and $c_{j}$ are clumped into a single cluster if

\begin{center}
$d\{c_{i},c_{j}\}=\min \{d\{c_{i},c_{j}\}\}$
\end{center}

with the distance between clusters being defined by

\begin{center}
$d\{c_{i},c_{j}\}=\min \{d_{pq}\}$ with $p\in c_{i}$ and $q\in c_{j}$
\end{center}

This process is continued until there is a single cluster. This clustering
process is also known as the \textit{single link method}, being the method
by which one obtains the minimal spanning tree (MST) of a graph.

In a connected graph, the MST is a tree of $N-1$ edges that minimizes the
sum of the edge distances. In a network with $N$ nodes, the hierarchical
clustering process takes $N-1$ steps to be completed, and uses, at each
step, a particular distance $d_{i,j}$ $\in $ $D$ to clump two clusters into
a single one.

Let $C=\{d_{q}\},q=1,...,N-1$, be the set of distances $d_{i,j}$ $\in $ $D$
used at each step of the clustering, and $L=\max \{d_{q}\}$. It follows that
$L=d_{N-1}$.

After computing the threshold distance value ($L$) we are able to define a
representation of $D$ with sparseness replacing fully-connectivity in a
suitable way.

The result of the hierarchical clustering process leading to the MST is
usually described by means of a dendrogram. During this process, a unique
color is assigned to each group of nodes within the dendrogram whose linkage
is less than $T$ times the value of the threshold distance $L$ ($\max
\{d_{q}\}$). In the dendrograms presented here, $T$ is set to $0.7$.

The dendrogram obtained from the matrix of distances presented in Figure 1
(a) is shown in the second plot (b) of Figure 1. There, three clusters can
be observed. The first cluster (colored blue) comprises the countries FI(8),
SE(21), DK(7), NO(18), NL(16), GB(23) and IT(13). The second cluster (black)
encompasses countries KY(6), FI(8), CH(22), AN(17), while the third (red)
comprises countries FR(9), DE(10), BE(4), LU(15), AT(1), CA(5), HK(11) and
JP(14). From this dendrogram we observe that the first two clusters seem to
be closely related, while a distinct situation characterizes the third one.
Countries BS (2) and BH(3) occupy quite distinct positions on the tree,
being the last ones to be connected in the hierarchical clustering process.
The third plot (c) in Figure 1 shows a network representation of the
corresponding MST. It is worth noting that two countries are close
(short-distant) on the MST whenever the evolution of their net claims is
correlated. Conversely, poorly correlated countries use to be those
occupying distant positions on the MST.

\begin{figure}[htb]
\begin{center}
\psfig{figure=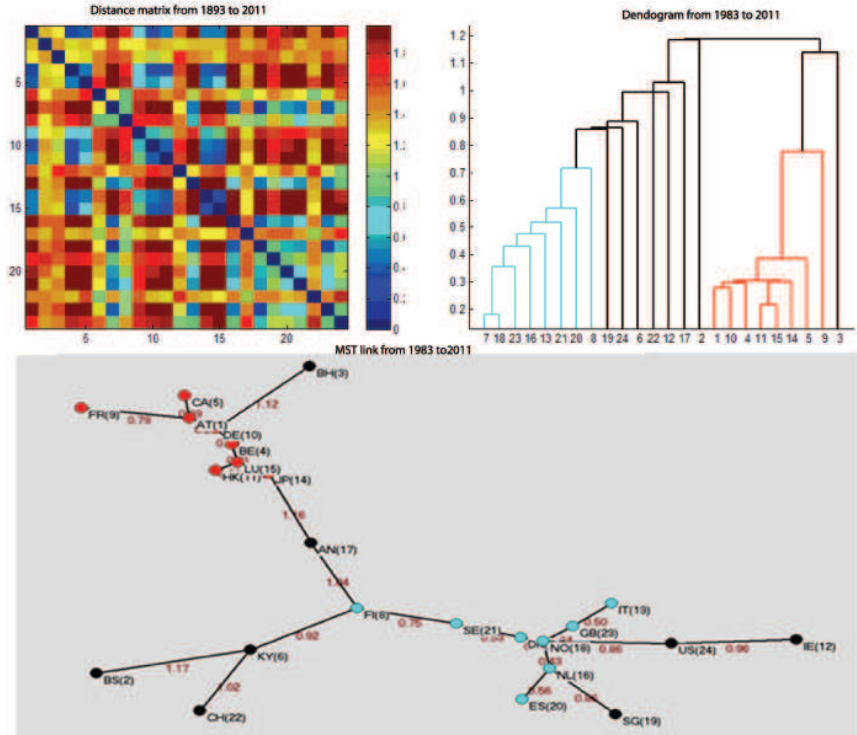,width=12truecm}
\end{center}
\caption{From 1983 to 2011 (a) the Matrix of Distances (b) the Dendrogram (c) the MST links}
\end{figure}

Plot (c) in Figure 1 highlights the role of Norway, which is connected to
several countries, displaying high centrality. The same happens to Austria
in the upper branch of this graph.

When a geographical perspective is taken, the first cluster may be
identified as the one representing the European peripherical countries,
including the northest countries FI(8), SE(21), DK(7), NO(18), NL(16),
GB(23) and two Mediterranean ones (IT(13) and ES(20)). The second cluster
brings together the central European countries FR(9), AT(1), DE(10), BE(4),
LU(15) but also includes the non-EU countries JP(14), CA(5) and HK(11) .
Such a geographically driven trend is in accordance with the recent study
presented in reference [12] on the geographical composition of countries'
external positions and its increasing trend since the mid-1990's.
Accordingly to this reference, this increasing trend in the countries'
positions has been specially pronounced among industrial countries, where
financial integrations has overstepped trade integration.

Besides the geographically-oriented nature, the MST representation of the 24
reporting countries does not seems to reveal other relevant concerns.

\subsection{Looking at chronologically successive periods}

To have a comparative idea on the structure inferred from the BIS data, we
have divided the data in two chronologically successive batches - from 1983
to 1997 and from 1997 to 2011 - and performed the same operations.

\begin{figure}[htb]
\begin{center}
\psfig{figure=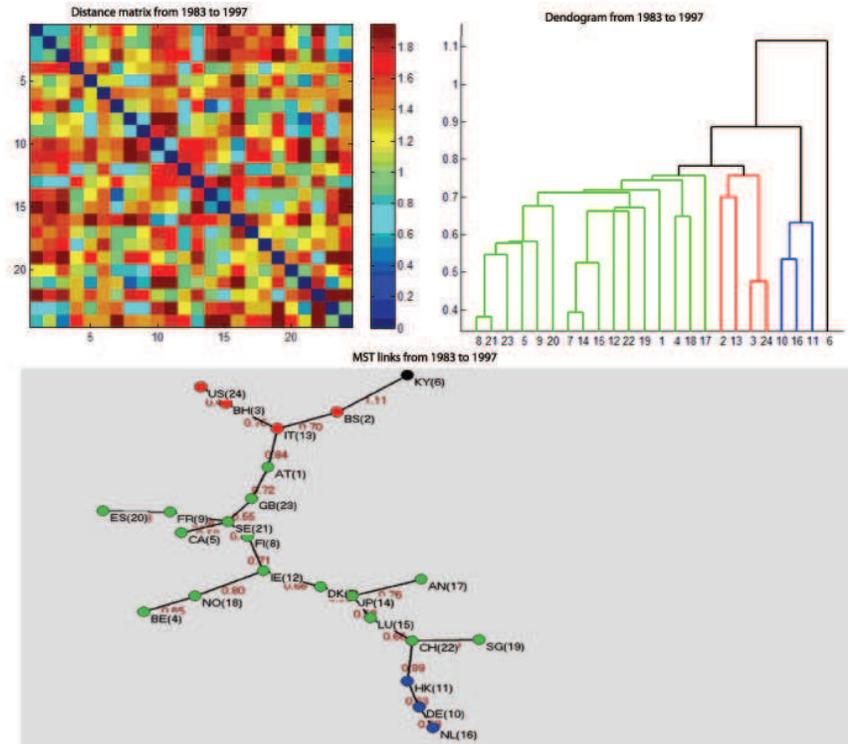,width=12truecm}
\end{center}
\caption{From 1983 to 1997 (a) the Matrix of Distances (b) the Dendrogram (c) the MST links}
\end{figure}

The three plots in Figure 2 show the distance matrix, the dendrogram and the
network representation of the MST obtained for the period 1983-1997.

The first cluster (green) is the biggest one and encompasses countries
AT(1), GB(23), SE(21), CA(5), FR(9), ES(20), FI(8), IE(12), NO(18), BE(4),
DK(7), JP(14), AN(17), LU(15), CH(22) and SG(19). The second (blue), being
highly correlated with the first, is comprised of countries HK(11), DE(10)
and NL(16). The third cluster (red) includes countries IT(13), BS(2), BH(3)
and US(24). Country KY(6) is the latest one to be connected on the MST. It
is worth noting that the need to include this country in the MST structure
introduces an increase in the threshold distance value ($L$) of more than
20\%.

Now clusters are no longer geographically-oriented and the dendrogram seems
to be characterized by greater uniformity in the distances between
countries. These distances are shortened, leading the whole set of countries
to apparently converge to a single cluster.

\begin{figure}[htb]
\begin{center}
\psfig{figure=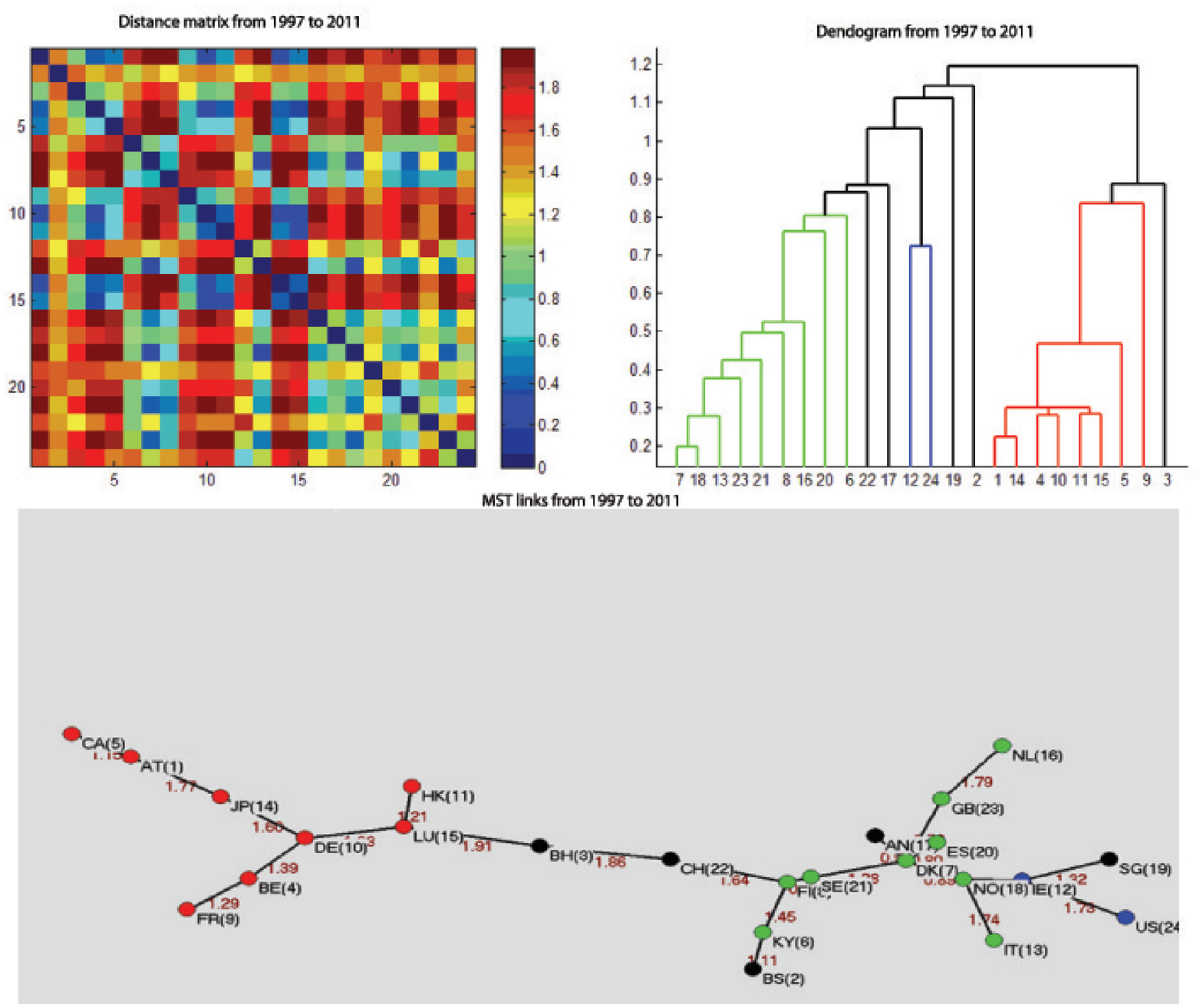,width=12truecm}
\end{center}
\caption{From 1997 to 2011 (a) the Matrix of Distances (b) the Dendrogram (c) the MST links}
\end{figure}

The three plots in Figure 3 show the distance matrix, the dendrogram and the
MST obtained for the period 1997-2011, where structural changes have led to
important modifications in the corresponding MST, mostly when compared with
the results obtained for the period 1983-1997.

Three clusters are present, the first one (green) includes countries KY(6),
FI(8), SE(21), DK(7), NO(18), IT(13), ES(20), GB(23) and NL(16). The second
(blue) is formed by only two countries IE(12) and US(24), while the third
(red) encompasses countries CA(5), AT(1), JP(14), DE(10), BE(4), FR(9),
LU(15) and HK(11). The corresponding MST informs that the two main clusters
are linked together by the two farthest countries (BH(3) and CH(22)) among
the whole set. The emergence of long distant links (low intensity
connections) playing such a central role use to be associated to less robust
network structures, whenever robustness is defined as propensity to
connectedness preservation.

Now, Denmanrk (DK(7)) and Germany (DE(10)) seem to be the countries with
greater centrality. The threshold distance value increases in this last
period, highlighting a higher degree of separation between the two largest
clusters.

It shall be noticed that these first results suggest that some structural
changes taking place in this last period are better observed when countries
are classified accordingly their debtor or creditor roles. In so doing, the
MST representation in Figure 3 shows a remarkable concentration of heavily
indebted countries in the first cluster, leaving the third cluster to be
comprised of countries that play the creditor role.

Structural changes in this period were certainly driven by the occurrence of
some stress events as the Asian (1997, 2nd Black Monday), Russian (1998) and
Japanese (1999) crises in the latest nineties. Instead of a
geographically-oriented structure, the network emerging in this last period
seems to be driven by the debtor/creditor roles.

\section{Beyond the MST approach}

In the last section, the construction of a MST allowed for the definition of
networks where sparseness replaces fully-connectivity in a suitable way.
However this construction neglects part of the information contained in the
distance matrix, since it only takes the $N-1$ distances that are considered
in the hierarchical clustering process.

In order to avoid this loss of information we define the projected graph $B$
(with $N$ vertices being the network nodes) by setting $b_{i,j}$ $=d_{i,j}$
if $d_{i,j}<$ $L$ and $b_{i,j}=0$ if $d_{i,j}$ $>L$. As usual, null arcs of $%
B$ are those for which $b_{i,j}=0$. Here we want to consider two nodes $i$
and $j$ to be connected if $d_{i,j}<$ $L$.

Let $A$ be the boolean graph associated with $B$, where each element $%
a_{i,j} $ is the number of edges of $B$ that join the vertices $i$ and $j$
and, since $B$ is a simple graph, $a$ $\in $$\{0,1\}.$

The three plots in Figure 4 show the boolean graphs ($A$) obtained for the
three periods: 1983-2011, 1983-1997, 1997-2011. They were obtained by:

\begin{enumerate}
\item Taking the matrix of distances ($D$) of each period (presented in the
first plots of Figures 1, 2 and 3).

\item Applying the hierarchical clustering process to obtain the distance $%
d_{N-1}\in C$ used in the last step of the hierarchical clustering process.

\item Building the corresponding boolean graph ($A$) where unit arcs ($%
d_{ij}\leq d_{N-1}$) are represented as black patches and null arcs ($%
d_{ij}>d_{N-1}$) correspond to white ones.
\end{enumerate}

\begin{figure}[htb]
\begin{center}
\psfig{figure=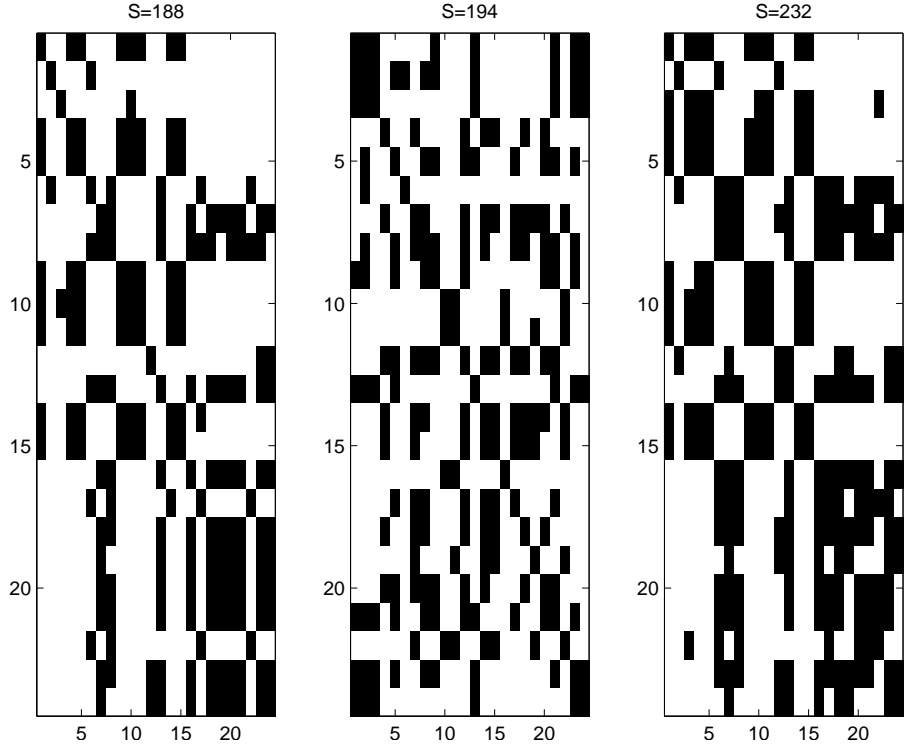,width=12truecm}
\end{center}
\caption{The boolean graphs (a) 1983-2011 (b) 1983-1997 (c) 1997-2011}
\end{figure}

Having defined the boolean representation associated to each time period, we
are also interested in defining $C^{\ast }=\{d_{l}\},l=1,...,M$ as the set
of distances in $D$ whose values are less or equal to $L$ and computing $%
S=M-(N-1)$.

Clearly $S>0$ is the number of \textbf{redundant} elements in $C^{\ast }$,
that is, the number of distances $d_{i,j}$ that, although being smaller than
$L$, need not be considered in the hierarchical clustering process.

In a connected graph, $S$ provides the cardinality of the cycle basis of the
graph, or its \textit{cyclomatic number}. Being a cycle basis of a graph
defined by the set of its elementary cycles that taken together yield the
entire graph. In this context, cycles and trees (i.e., connected graphs
without cycles) may or not appear in the resulting network structures [21].
The clustered networks have a high coordination number while in the opposite
case the networks approach a tree-like structure and, consequently, low
clustering.

Computing $S$ for the three time periods 1983-2011, 1983-1997 and 1997-2011
yields respectively $188$, $194$ and $232$. The value of $S$ obtained for
the later period shows an increase in the number of redundant links. Not
surprisingly, increasing cluster use to be a consequence of disturbed
periods being the topological correlate of the occurrence of stress events.

It has been empirically observed that both during economic expansion and
normal periods, financial markets tend toward randomness whereas in the
disturbed periods its structure is reinforced in the topological sense, as
revealed by the clustering measures (a detailed discussion on the role of
clustering in financial crises can be found in references [22], [23] and
[24]). The number of redundant elements that characterize the distinct time
periods provides enough evidence on the structural changes taking place on
the network structure. They are due to the emergence of high correlated
positions (synchronization) in the network of countries.

To better capture structural changes we also define the \textbf{residuality }%
coefficient

\begin{equation}
R=\frac{\sum_{d_{i,j}>L}d_{i,j}^{-1}}{\sum_{d_{i,j} \le L}d_{i,j}^{-1}}
\end{equation}

where $L$ is the highest threshold distance value that insures connectivity
of the whole network in the hierarchical clustering process.

\begin{figure}[htb]
\begin{center}
\psfig{figure=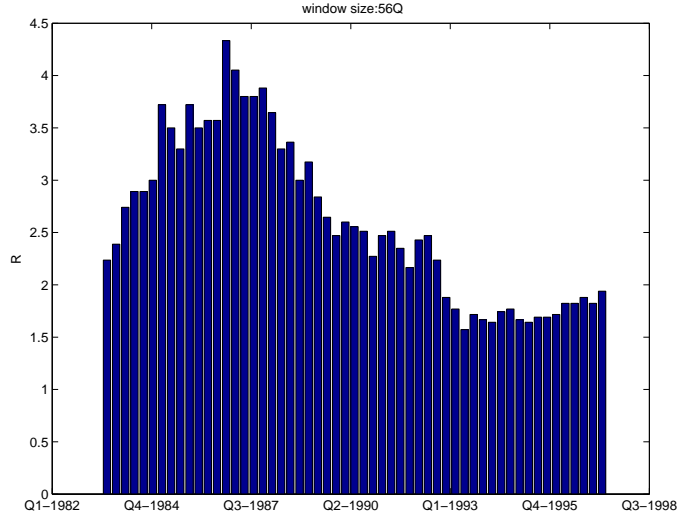,width=9truecm}
\end{center}
\caption{The coefficient R (1983-2011) with a time window of 56 quarters}
\end{figure}

Residuality relates the relative strengths of the connections above and
below the threshold distance value.

Figure 5 shows the values of $R$ along the 28 years and with a time window
of 56 quarters (14 years). At each quarter ($t$) $R$ is computed from the
distances measured along a moving window of length 56. The choice of such a
wide window ensures statistical robustness and allows for capturing the main
differences in the behavior of $R$ when the latest nineties start to be
taken into account. Indeed, Figure 5 shows that the lowest values of $R$
correspond to the 14 years starting at the last quarter of 1993 and that,
after the end of 1997, $R$ does not return to the previous configuration.

The decrease of the residuality coefficient observed in Figure 5 seems to be
twofold: it happens because the connection strengths $d_{ij}^{-1}$ above $%
\frac{1}{L}$ tend to be stronger than those that remain below the threshold.
In other words, the distance values below $L$ tend to be smaller (shorter)
than those above the threshold, showing that synchronization between
countries' positions increases significantly. Moreover, the decrease of $R$
is due to the fact that\ the networks become less sparse (the number of
links increases) after 1997, forcing several \textit{residual links} to
leave this category.

As partially revealed by the increase of $S$, the values of the overall
distances decrease after 1997 but there is an increase of the distance
between the two lagest clusters, as well a corresponding increase of the
value of the endogenous threshold $L$. The results reveal that a major
change is occurring in the last period. This difference in the empirically
described evolution suggests that in the period of the easy interest rates,
the Internet boom and the housing bubble, a new network regime was generated
from the cross-border financial linkages.

As Milesi-Ferretti and co-authors point out [25], the last 15 years were
characterized by an explosion in the size of cross-border capital flows,
measured by the rise of exchanges of claims and liabilities. Consequently,
these years have been faced with the emergence of global imbalances and a
sharp widening of debtor and creditor positions. The behavior of $R$, while
capturing the relative strengths of the links above and below an endogenous
threshold, helps to characterize the impact of the global crises on the
structure of the networks of cross-border capital flows.

In the next section, the classification into debtor and creditor countries
improves the empirical characterization of the cross-border and historically
dependent networks.

\section{Creditor and debtor networks}

Now we are interest in the average position $\bar{p}_{i}$ of each country
computed for each of the three periods (1983-2011, 1983-1997, 1997-2011)
considered so far. The average position of country $i$ is given by

\begin{equation}
\bar{p}_{i}=\left\langle C_{i,t}\right\rangle -\left\langle
L_{i,t}\right\rangle =\frac{\sum^{t=n}(C_{i,t}-L_{i,t})}{n}
\end{equation}

where $n$ is the number of time labels. In the case of the three periods
under analysis, $n$ is set to 110, 55 and 55 respectively.

From this notion of \emph{average position }- which corresponds to the
average net claims of each country - we are able to classify countries as
creditor or debtor countries accordingly to the sign of $\bar{p}_{i}$. Table
2 summarizes the average position of each country computed for the three
time intervals.

\begin{small}
\begin{center}
\begin{tabular}{|c|c|c|c|c|c|c|c|}
\hline\hline
\emph{Country} & 83-11 & 83-97 & 97-11 & \emph{Country} & 83-11 & 83-97 &
97-11 \\ \hline\hline
Austria & C & D & C & Italy & D & D & D \\ \hline\hline
Bahamas & D & D & D & Japan & C & C & C \\ \hline\hline
Bahrain & C & C & C & Luxemburg & C & C & C \\ \hline\hline
Belgium & C & D & C & Netherland & D & C & D \\ \hline\hline
Canada & D & D & C & Netherland Antilles & C & C & C \\ \hline\hline
Cayman Islands & C & C & C & Norway & D & D & D \\ \hline\hline
Denmark & D & C & D & Singapore & D & C & D \\ \hline\hline
Finland & D & D & D & Spain & D & D & D \\ \hline\hline
France & C & D & C & Sweden & D & D & D \\ \hline\hline
Germany & C & C & C & Switzerland & C & C & C \\ \hline\hline
Hong Kong & C & C & C & United Kingdom & D & D & D \\ \hline\hline
Ireland & D & D & D & United States & D & D & D \\ \hline\hline
\end{tabular}

{\small Table 2: Creditor (C) and Debtor (D) countries}
\end{center}
\end{small}

Once countries are assigned to their debtor or creditor roles, the
dendrograms and corresponding MST networks are presented in Figures 6, 7 and
8. Creditor countries are colored green while debtor ones are colored red.

\bigskip The observation of these figures gives rise to the following
remarks:

\begin{enumerate}
\item Debtor and creditor countries occupy (were placed in) separated areas,
both in the dendrograms and in the MST networks, showing a clear
segmentation of the whole set of countries into two clusters.

\item The average distance value inside each class is significantly smaller
than the average network distance.

\item Some networks, as in Figure 6, allows for the observation of both a
core group and a peripherical group of countries, whose main role seems to
be to tie together small groups of debtor and creditor ones.

\item There is an important increase in the threshold distance value in the
last 14 years, since $L_{83-97}=3.94$ and $L_{97-11}=5.52$ .

\item There is a corresponding increase in the intercluster distance value
in the later period, since it increases from $0.66$ (between DK(7) and
IE(12)) to $0.86$ (between FI(8) and CH(22)).

\item Some countries remain not affected in their relative average position
on the network structure, they are: BS(2) and CA(5) during the period
1983-2011, KY(6) and BH(3) in 1983-1997, and KY(6) and AN(17) in the later
period. This result seem to be due to the existence of a typical trend in
the sign of their net claims. Countries KY(6) and AN(17) are both creditors
but their performance during the last period is worsening. The same happens
with countries BH(3) and KY(6) in the period 1983-1997, while the
performances of both countries BS(2) and CA(3) show an improving trend even
if these countries occupy a average debtor position along 1983-2011.
\end{enumerate}

\begin{figure}[htb]
\begin{center}
\psfig{figure=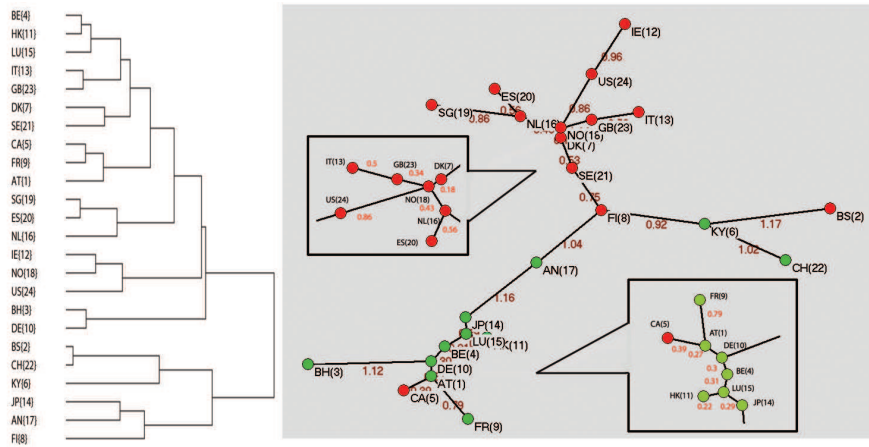,width=12truecm}
\end{center}
\caption{The Dendrogram and the MST Links for the period 1983-2011}
\end{figure}

When debtor and creditor positions are computed for the shorter time periods
(1983-1997 and 1997-2011) an emphasis should be placed on the size of the
corresponding debts or credits. To this end, Figures 7, 8 present the MST
and the dendrogram obtained in each chronological period using large circles
to represent both the largest creditors and the largest indebted countries.
Results show that:

\begin{enumerate}
\item In the first period (Figure 7), eleven countries give rise to a
cluster containing the largest indebted countries, which is linked to a
second cluster comprising the largest creditor ones.

\item In the second period and accordingly to the dendrogram in Figure 8,
the splitting into two separated clusters is less clear. However the network
representation of the corresponding MST shows that the network segmentation
into creditor and debtor branches remains.

\item In both periods, results show that the three main indebted countries
are IT(13), GB(23) and US(24).

\item The branch with the largest creditors is less stable, comprising
countries JP(14), DE(10) and CH(22) in the first period and being later
comprised of countries JP(14), DE(10) and LU(15).
\end{enumerate}

In general terms, the first period is characterized by a less pronounced
splitting into debtor and creditor countries, being two of heavily indebted
countries directly linked to creditor ones. Conversely, after the latest
nineties, the heavily creditor countries give rise to a cluster not directly
connected to the debtor ones. In what concerns the heavily indebted
countries, this last and critical period reinforced their separation from
the creditor ones.

\begin{figure}[htb]
\begin{center}
\psfig{figure=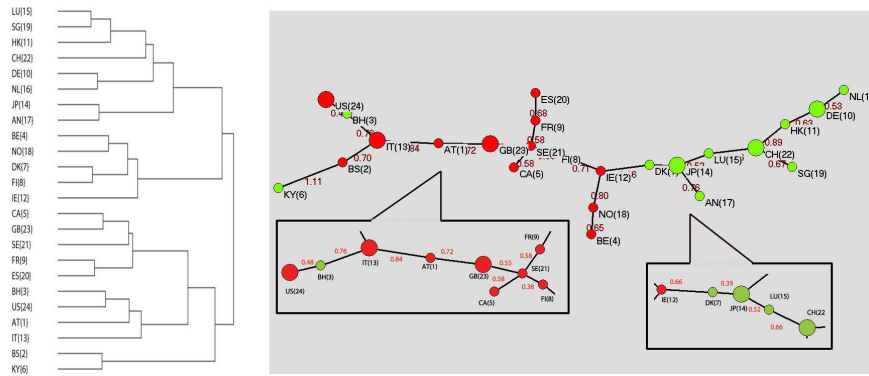,width=12truecm}
\end{center}
\caption{The Dendrogram and the MST Links for the period 1983-1997}
\end{figure}

\begin{figure}[htb]
\begin{center}
\psfig{figure=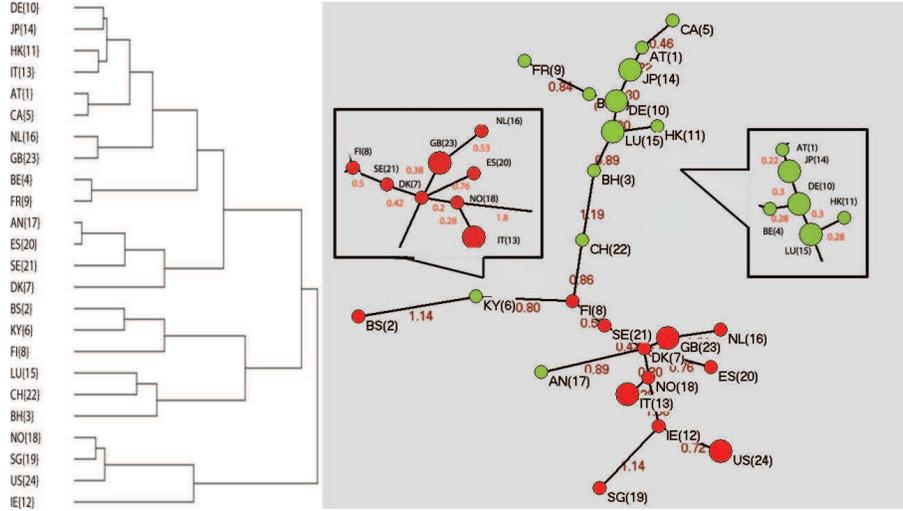,width=12truecm}
\end{center}
\caption{The Dendrogram and the MST Links for the period 1997-2011}
\end{figure}

\section{Concluding Remarks}

Using time series of interbank liabilities and claims conducted through the
international banking system, we have developed complete networks of
positions (net claims) between 24 countries. These structures were developed
for \textit{(i)} a 28-years period (1983-2011) and\emph{\ }two 14-years
periods \textit{(ii)} (1983-1997) and\textit{\ (iii)}\emph{\ }(1997-2011).

From the fully-connected representations of \textit{(i)}, \textit{(ii)} and%
\textit{\ (iii)}, the sparse and connected networks of financial linkages
across countries were constructed through the Minimal Spanning Tree
approach. To reduce the loss of information resulting from the MST
construction we have accounted for the number and the intensity of the
redundant linkages in each time period. A coefficient of residuality is thus
defined to capture the structural changes occurring on the network along the
last 28 years. This coefficient highlights an important modification acting
in the financial linkages in the period 1997-2011, and situates the
turbulence taking place at the global financial system since the Summer 2007
as replica of a larger structural change going on for a decade.

We have also addressed the role of debtor and creditor countries in the
networks obtained from \textit{(i)}, \textit{(ii)} and\textit{\ (iii)}
branches of data. Results showed that the period after 1997 is characterized
by a stronger splitting of the network into debtor and creditor countries.
Such a splitting seems to be even strengthened when emphasis is placed on
the size of the corresponding credits and debts.

As the data and the method suggest, this ongoing turbulence is part of a
mutation in the structure of the cross-border interdependencies since the
latest nineties. Results highlight an important modification acting in the
financial linkages across countries in the period 1997-2011, and situate the
recent financial crises as replica of a larger structural change going on
since 1997.

In future work we envision the application of the same approach to the
analysis of networks of liabilities. By placing emphasis on the relative
amounts of debts we envision the development of a deeper topological
description of the cross-border financial linkages, still beyond the MST
approach.

\bigskip

\textbf{Acknowledgement}: \emph{This work has benefited from partial
financial support from the Funda\c{c}\~{a}o para a Ci\^{e}ncia e a
Tecnologia-FCT, under the 13 Multi-annual Funding Project of UECE, ISEG,
Technical University of Lisbon.}

\end{document}